\documentclass[12pt,preprint]{aastex}

\newcommand{\bvf}{Brunt-V\"{a}is\"{a}l\"{a} }
\newcommand{\teffm}{$\rm T_{\rm eff}$}
\newcommand{\teff}{\rm T_{eff}}
\newcommand{\msunm}{$\rm M_\odot$}
\newcommand{\msun}{\rm M_\odot}

\newcommand{\pdots}{\.{P}'s}
\newcommand{\deltap}{\langle \Delta P \rangle}
\newcommand{\deltapm}{$\langle \Delta P \rangle$ }
\newcommand{\massteff}{$\rm M_* - T_{eff}$ }

\shorttitle{Asteroseismology of R548 and G117-B15A}
\shortauthors{Bischoff-Kim et al.}

\title{Fine Grid Asteroseismology of G117-B15A and R548}
\author{A. Bischoff-Kim, M.H. Montgomery, D.E. Winget}
\affil{The University of Texas at Austin, Astronomy Department, 1 University Station, C1400, Austin, TX 78712, USA}
\email{agnes@astro.as.utexas.edu, mikemon@astro.as.utexas.edu, dew@astro.as.utexas.edu}

\begin{abstract}
We now have a good measurement of the cooling rate of G117-B15A. In the near 
future, we will have equally well determined cooling rates for other pulsating 
white dwarfs, including R548. The ability to measure their cooling rates offers
us a unique way to study weakly interacting particles that would contribute to 
their cooling. Working toward that goal, we perform a careful 
asteroseismological analysis of G117-B15A and R548. We study them side by side 
because they have similar observed properties. We carry out a systematic, fine 
grid search for best fit models to the observed period spectra of those stars. 
We freely vary 4 parameters: the effective temperature, the stellar mass, the 
helium layer mass, and the hydrogen layer mass. We identify and quantify a 
number of uncertainties associated with our models. Based on the results of 
that analysis and fits to the periods observed in R548 and G117-B15A, we 
clearly define the regions of the 4 dimensional parameter space ocuppied by the
best fit models.
\end{abstract}


\begin{document}

\section{Astrophysical context}

G117-B15A and R548 are pulsating white dwarfs with atmospheres dominated by 
hydrogen. These stars are called DAVs or ZZ Ceti stars. Their effective 
temperatures are close to 12,000 K. They are non-radial, g-mode pulsators, 
where the restoring force is buoyancy. Because of its potential use to study 
axions and other exotic physics \citep{isern, Kepler2}, G117-B15A was the 
object of several asteroseismological studies. R548 is similar to G117-B15A 
both spectroscopically and pulsationally so studying them side by side may 
provide additional clues. \citet{paul98} performed such a parallel study. More 
recently \citet{benvenuto} published an asteroseismological fit to G117-B15A's 
period spectrum. \citeauthor{benvenuto} evolved models from the Main Sequence 
and included detailed physics, such as time-dependent diffusion. Their best fit
model has an effective temperature and a mass consistent with spectroscopic 
results, but the periods of the model do not match the observed period spectrum
as well as \citeauthor{paul98}'s fits do.

We take a new approach to white dwarf asteroseismology, enabled by the 
computing resources available today. Using models similar to 
\citeauthor{paul98}'s, we perform a systematic, fine grid search for the best 
fits to the period spectra of G117-B15A and R548. We study both stars side by
side. While we do not ignore the spectroscopy entirely, we take a naive 
approach and base much of our results on the period fitting alone. At the end 
we compare our results with the spectroscopy and previous studies. In the same 
naive spirit, we treat the helium layer mass as another free parameter. 
\citet{paul98} and \citet{benvenuto} both fixed its value at $10^{-2} M_*$, 
where $M_*$ is the total mass of the model.

We organize our paper as follows. In section 2, we present the observables we 
have for G117-B15A and R548. In section 3, we summarize concisely the 
asteroseismological results of \citet{benvenuto} and \citet{paul98}. In section
4, we explain our method. We include a discussion of our models and a 
quantitative study of what input parameters matter most. We present our results
in section 5, along with a discussion of how they compare with the fits done by
\citet{paul98} and \citet{benvenuto}. We summarize and discuss our results in 
section 6.

\section{The stars}

G117-B15A and R548 have much in common and as we shall see, studying them
side by side helps constrain the model parameters. In Table 
\ref{starsproperties} we list the spectroscopically determined effective 
temperatures and gravities of both stars, along with the modes we observe in 
those stars \citep{Anjum1}. These are the clues we have at our disposal to find
best fit models to G117-B15A and R548. 

\begin{table}[!ht]
\caption{
\label{starsproperties}
Observed properties of G117-B15A and R548}
\begin{center}
\begin{tabular}{lll}
\noalign{\smallskip}
\hline
\hline
\rule[-0.3cm]{0mm}{0.8cm}
& G117-B15A & R548\\
\hline
\rule[-0.0cm]{0mm}{0.6cm}
$\teff$            & $11620 \pm 200$ (a)             &  $11990 \pm 200$ (a) \\
                   & $12830 \pm 350$ (b)             &                      \\
                   & $12250 \pm 125$ (c)             &                      \\
                   & $12010 \pm 180$ (d)             &  $11865 \pm 170$ (d) \\
\rule[-0.3cm]{0mm}{0.6cm}
                   & $12375 \pm 125$ (e)             &                      \\
\hline
\rule[-0.0cm]{0mm}{0.6cm}
logg               & $7.97 \pm 0.05$ (a)             &                      \\
                   & $7.92 \pm 0.05$ (b)             & $7.97 \pm 0.05$ (b)  \\
                   & $8.10 \pm 0.15$ (c)             &                      \\
\rule[-0.3cm]{0mm}{0.6cm}
                   & $7.94 \pm 0.17$ (d)             & $7.89 \pm 0.17$ (d)  \\
\hline
\rule[-0.0cm]{0mm}{0.6cm}
Periods [s] )      & 215.22 (19.8)             & 212.77 (4.1), 213.13 (6.7) \\
(Amplitudes [mma]) & 270.86 (7.1)              & 274.25 (4.1), 274.78 (2.9) \\
                   & 304.15 (8.8)              &                            \\
                   &                           & 187.27 (1.0)               \\
                   &                           & 318.08 (0.9)               \\
\rule[-0.3cm]{0mm}{0.6cm}
                   &                           & 333.65 (1.0)               \\
\hline
\end{tabular}
\end{center}
\tablerefs
{(a) \citet{b95b}; (b) \citet{b95a}; (c) \citet{koester94};
(d) \citet{koester1}; (e) \citet{elr95}.}
\end{table}

The most solid observables listed in Table \ref{starsproperties} are the 
periods. For both G117-B15A and R548, periods are determined to better than a
second. The 215s mode in G117-B15A and the 213s in R548 are very stable
and determined to extremely high accuracy. By comparing the amplitude of 
G117-B15A's 215s mode in G117-B15A in the ultra violet to those in the visible,
\citet{elr95} were able to determine confidently that it was an $\ell=1$ mode.
R548's higher amplitude modes are close to the 215s and 271s modes in G117-B15A
and are likely $\ell=1$ modes as well. 

Effective temperature and gravity are more poorly determined than the periods. 
We give an overview of the current spectroscopy in Table \ref{starsproperties}.
To obtain an effective temperature for G117-B15A, \citet{elr95} assumed a log g
of 7.97, following work done by \citet{daou}.

\section{Previous fits to G117-B15A and R548}

We summarize the best fit models \citet{benvenuto} found for G117-B15A in
Table \ref{g117benvenuto} and those \citet{paul98} found in Table 
\ref{g117paul}. $M_H$ is the hydrogen layer mass and $M_{He}$ the helium layer
mass. We define $\Phi$ in equation \ref{functominimize}. In essence,
it is the average difference between the observed periods and the 
calculated periods. The lower $\Phi$, the better the fit.

\begin{table}[!ht]
\caption{
\label{g117benvenuto}
Best fits to G117-B15A's observed period spectrum according to 
\citet{benvenuto}.}
\begin{center}
\begin{tabular}{lll}
\noalign{\smallskip}
\hline
\hline
\rule[-0.3cm]{0mm}{0.8cm}
Mode identification & $k=1,2,3$ & *$k=2,3,4$             \\
\hline
\rule[-0.0cm]{0mm}{0.6cm}
$\teff$ [K]         & 11400     & 11800                  \\
~$\rm M_*/\msun$     & 0.500     & 0.525                 \\
~$\rm M_H/M_*$       & $10^{-6}$ & $1.48 \times 10^{-4}$ \\
~$\rm M_{He}/M_*$    & $10^{-2}$ & $10^{-2}$             \\
\rule[-0.3cm]{0mm}{0.6cm}
$\Phi$ [s]          &           & 6.5                    \\
\hline
\hline
\end{tabular}
\end{center}
\tablecomments
{* \citeauthor{benvenuto} picked this fit because it matched the spectroscopy 
of \citet{koester00} better. \citeauthor{benvenuto} do not give a list of 
calculated periods for the $k=1,2,3$ fit, but they show and state that it is 
comparable in quality to the $k=2,3,4$ fit.}
\end{table}

\begin{table}[!ht]
\caption{
\label{g117paul}
Best fits to G117-B15A's observed period spectrum according to \citet{paul98}}
\begin{center}
\begin{tabular}{l|lll|lll}
\noalign{\smallskip}
\hline
\hline
\rule[-0.3cm]{0mm}{0.8cm}
Mode identification &\multicolumn{3}{c|}{$k=1,2,3$}
                                              &\multicolumn{3}{c}{$k=2,3,4$}\\
\hline
\rule[-0.0cm]{0mm}{0.6cm}
$\teff$ [K]        & 12160  & 11460  & 10790   & 12530  & 11860  & 11120     \\
~$\rm M_*/\msun$   & 0.55   & 0.60   & 0.65    & 0.55   & 0.60   & 0.65      \\
~$\rm M_H/M_*$     & $3 \times 10^{-7}$  &       
                        $2 \times 10^{-7}$  &       
                           $1 \times 10^{-7}$  & $1.5 \times 10^{-4}$  &
                                                    $8 \times 10^{-5}$  &
                                                       $1 \times 10^{-5}$    \\
~$\rm M_{He}/M_*$  &\multicolumn{3}{c|}{$10^{-2}$}
                                              &\multicolumn{3}{c}{$10^{-2}$}\\
\rule[-0.3cm]{0mm}{0.6cm}
$\Phi$ [s]          & 0.5   & 0.8    & 1.3     & 1.1    & 1.3    & 2.5       \\
\hline
\hline
\end{tabular}
\end{center}
\end{table}

Both authors find two families of solutions; one where the mode identification 
for G117-B15A's three observed periods is $k=1$, 2, and 3 and the other where 
it is $k=2$, 3, and 4 ($\ell=1$). \citeauthor{paul98} notes that for his 
models, both classes of solutions fit the observed periods and spectroscopic 
temperature equally well. Both authors find that the $k=1,2,3$ family of fits 
have thin hydrogen layers and the $k=2,3,4$ family thick hydrogen layers. 
\citeauthor{paul98}'s fits are hotter than \citeauthor{benvenuto}'s. Those are 
two trends to keep in mind.

Finally, we summarize the best fit models \citeauthor{paul98} found for R548 in
Table \ref{r548}. At the time, \citeauthor{paul98} had at his disposal only two
confirmed modes (the high amplitude doublets near 213s and 274.5s) and one 
likely mode (the 318s mode). Again, \citeauthor{paul98} finds two classes of 
solutions, one with the three known modes identified as $k=1,$ 2, and 3 and 
the other as $k=2$, 3, and 4 (all $\ell=1$).

\begin{table}[!ht]
\caption{
\label{r548}
Best fits to R548's observed period spectrum according to \citet{paul98}}
\begin{center}
\begin{tabular}{l|lll|lll}
\noalign{\smallskip}
\hline
\hline
\rule[-0.3cm]{0mm}{0.8cm}
Mode identification &\multicolumn{3}{c|}{$k=1,2,3$}
                                              &\multicolumn{3}{c}{$k=2,3,4$}\\
\hline
\rule[-0.0cm]{0mm}{0.6cm}
$\teff$ [K]        & 12440 & 11560  & 10790   & 12190  & 11970  & 11320     \\
~$\rm M_*/\msun$   & 0.55   & 0.60   & 0.65    & 0.55   & 0.60   & 0.65      \\
~$\rm M_H/M_*$     & $3 \times 10^{-7}$  &       
                        $1 \times 10^{-7}$  &       
                           $5 \times 10^{-8}$  & $1.5 \times 10^{-4}$  &
                                                    $5 \times 10^{-5}$  &
                                                       $5 \times 10^{-6}$    \\
~$\rm M_{He}/M_*$  &\multicolumn{3}{c|}{$10^{-2}$}
                                              &\multicolumn{3}{c}{$10^{-2}$}\\
\rule[-0.3cm]{0mm}{0.6cm}
$\Phi$ [s]          & 3.2   & 1.3    & 0.3     & 4.4   & 4.6    & 3.1       \\
\hline
\hline
\end{tabular}
\end{center}
\end{table}

In Table \ref{r548} we have re-evaluated $\Phi$ based on the knowledge that
the 318s mode is indeed real. \citeauthor{paul98} could not rely on that mode
and left it out. Including the 318s mode, we immediately learn something new: 
the $k=2,3,4$ fits are significantly worse than the $k=1,2,3$ fits. If we do 
not include the 318s mode, the two classes of fits are equally good. 
\citeauthor{paul98} concludes that the $k=2,3,4$ fits are better
because they agree with the spectroscopic temperatures better. He also computes
the rotational splitting of the two modes and finds that they also appear to 
favor the $k=2,3,4$ fits.

\section{Method}

In essence, the problem we have to solve in white dwarf asteroseismology is the
simple minimization of a function (the average difference between the 
calculated periods and the observed periods) with n variables, where n is 4 in 
the present study. Those variables include $\teff$, $\rm M_*$, $\rm M_{He}$ and
$\rm M_{H}$. Expressed mathematically:
\begin{equation}
\label{functominimize}
\Phi(\teff, M_*, M_{He}, M_{H}) \: = \: \deltap \: = \: \frac{A}{N}
\Sigma_{i=1}^{N} \; |P_i^{calc}-P_i^{obs}|
\end{equation}
where N is the number of observed periods and A is a normalizing
factor to account for the fact that at higher effective temperatures and 
higher masses, the period spacing decreases, artificially increasing our
chances of finding a good period match. For the regions of parameter space 
considered in this work, $A \sim 1$.

The simplest way to minimize $\Phi$ is to compute it for all conceivable values
of the 4 variables and pick the smallest value we find. But the number of times
we have to evaluate $\Phi$ can quickly become astronomical. To make matters 
worse, each evaluation of the function requires the full computation of a white
dwarf model. The White Dwarf Evolution Code (WDEC), described in the next 
section, allows us to compute a large number of models in a small amount of
time. But even with the WDEC, building a detailed map of $\Phi$ over all the 
relevant parameter space has only recently become practical on a standard 
desktop machine.

\subsection{Models}

The WDEC evolves hot polytrope models from temperatures close to 100,000K down
to the temperature of our choice. Models in the temperature range of interest 
for the present study are thermally relaxed solutions to the stellar structure 
equations. Each model we compute for the grid is the result of such an 
evolutionary sequence. 

The WDEC is described in detail in \citet{lvh75} and 
\citet{wood90}. We changed a few things since the work done by \citet{paul98}.
We used smoother core composition profiles and experimented with the more 
complex profiles that result from stellar evolution calculations 
\citep{salaris}. We updated the envelope equation of state tables from 
those calculated by \citet{oldeos} to those given by \citet{enveos} and use 
OPAL opacities \citep{opal}. We also treated the abundance of elements 
differently at the boundary between the helium layer and the hydrogen layer.
We assumed equilibrium diffusion profiles, following the prescription given
by \citet{arcoragi}, but do \emph{not} make the trace element approximation;
this was shown to yield results quite similar to those based on
time-dependent diffusion calculations \citep{althaus}. We still treat
diffusion at the carbon-helium transition zone as a free parameter.

\begin{figure}[!ht]
\begin{center}
\rotatebox{270}{\includegraphics[width=230pt,height=200pt,bb=25 200 580 600]{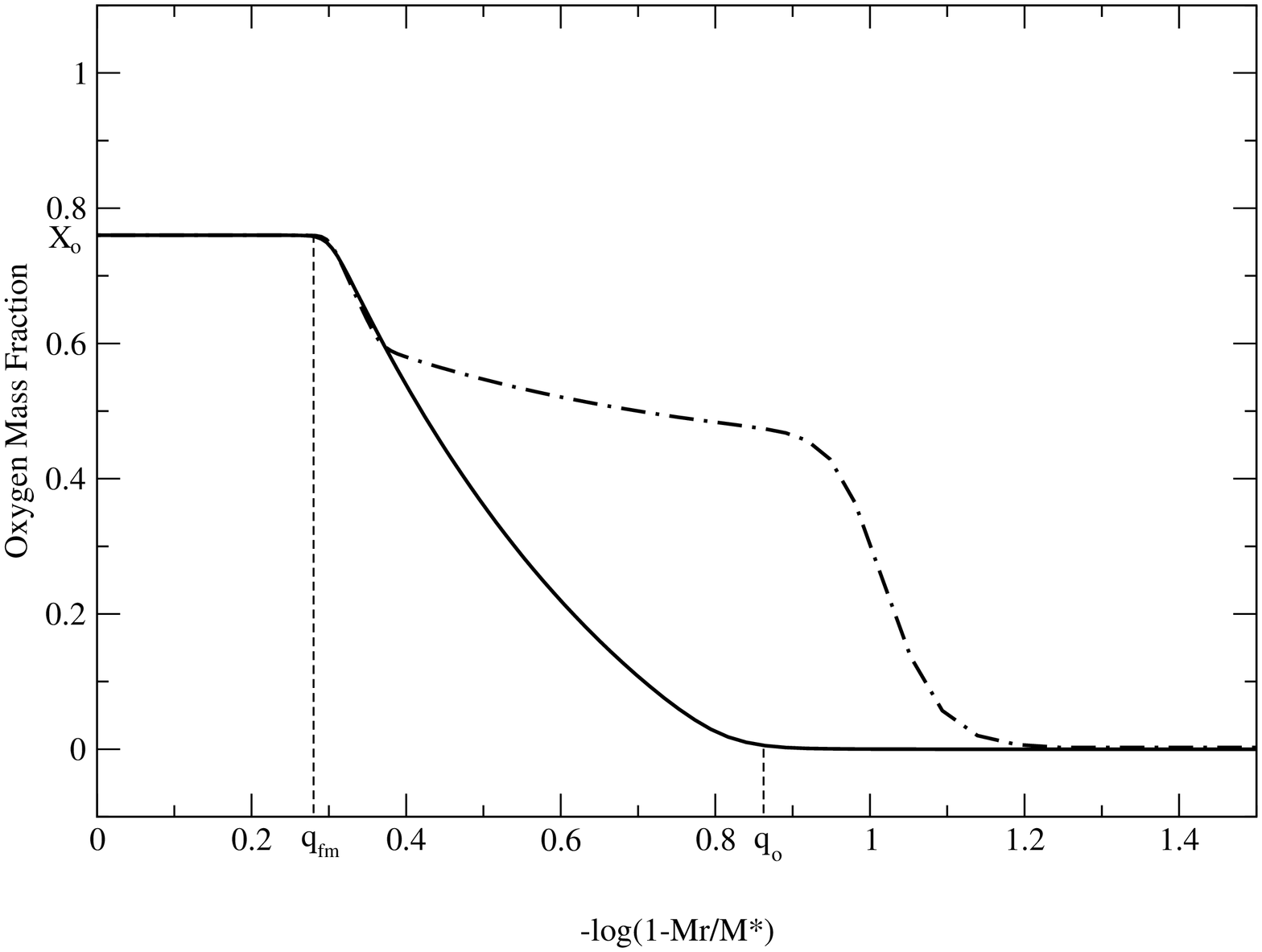}}
\end{center}
\caption{
\label{fig1}
Basic oxygen abundance profile (solid curve) and the three associated 
parameters $\rm X_o$, $\rm X_{fm}$, and $\rm q_o$. The dashed-dotted curve is 
an example of a Salaris-like oxygen abundance profile.
}
\end{figure}

\begin{figure}[!ht]
\begin{center}
\rotatebox{270}{\includegraphics[width=230pt,height=200pt,bb=25 200 580 600]{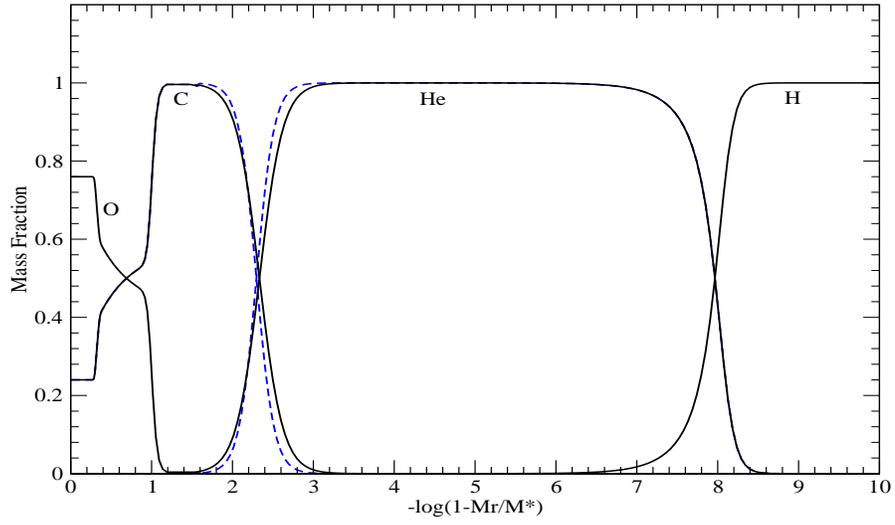}}
\end{center}
\caption{
\label{fig2}
Chemical composition profiles for the fiducial model (solid lines) and for a 
model with a sharper carbon-helium transition zone (dashed lines).
}
\end{figure}

There are three parameters associated with the shapes of the oxygen (and 
carbon) core composition profiles used by \citet{paul98}: the central oxygen
abundance ($\rm X_o$), the edge of the homogeneous carbon and oxygen core 
($\rm q_{fm}$), and the point where the oxygen abundance drops down to zero
($\rm q_o$). q is a mass coordinate, defined as $q(r) = -\log(1 - M(r)/M_*)$, 
where M(r) is the mass enclosed in radius r and $M_*$ is the stellar mass. We 
show an example of a basic oxygen abundance profile and those three parameters 
in figure \ref{fig1} along with a Salaris-like profile. In figure \ref{fig2}, 
we show abundance profiles for a fiducial model (12,400K, 0.62 \msunm, 
$\rm M_{He} = 10^{-2.3} M_*$, and $\rm M_{H} = 10^{-7.7} M_*$), along with
a model with a sharper carbon-helium transition zone. We shall use both
models in the next section. From now on, the $\rm M_*$ in the expression of the
helium and hydrogen layer masses will be implicit. We adopted Salaris-like 
profiles for the carbon and oxygen abundances. 

\subsection{The Parameters that Matter Most}

In order to determine the relative importance of each factor, we computed  
periods for the fiducial model we described above (one of our best fit models 
for G117-B15A) and then varied one parameter at a time. Each time, we compared 
the resulting periods with the period spectrum of the fiducial model. In our 
analysis, we did not compare the full calculated period spectra, but instead 
picked the three periods that were closest to those found in the observed 
period spectrum of G117-B15A. This gives us an idea of what parameters matter 
most for fitting G117B15A and R548 in particular (though the same parameters in
some order also matter for asteroseismological fits to other white dwarfs).  

We list the results in order of decreasing significance in Table 
\ref{uncertainties}. In column 1, we list the properties of the fiducial 
model, in column 2 the changes we made in each case, and in column 3 the 
average difference between the periods of the fiducial model and the modified 
model (equation \ref{functominimize}, with A = 1). We also detail the 
difference for each of the three periods of the fiducial model that match 
G117-B15A's observed period spectrum (216.3s, 270.9s, and 304.6s). The last two
lines in Table \ref{uncertainties} refer to the MLT treatment of convection in 
the models. $\alpha$ is the ratio of the pressure scale height to the mixing 
length. ML2 refers to the prescription of \citet{ml2} and ML1 to that of 
\citet{ml1}. While convection is responsible for important non-linear effects 
in the light curves, \citep[e.g.][]{mike05}, it has very little effect on the 
periods of the modes excited in the models because it only affects the outer 
$10^{-12}$ mass fraction of the models or less.

\begin{table}[!ht]
\caption{
\label{uncertainties}
Parameters that matter most for DAV models}
\smallskip
\begin{center}
\begin{tabular}{llllll}
\hline
\hline
\rule[-0.3cm]{0mm}{0.8cm}
Fiducial model  & Change                       & \deltapm  
           & $\Delta P_{216s}$  & $\Delta P_{271s}$ & $\Delta P_{305s}$     \\
\hline
\rule[-0.0cm]{0mm}{0.6cm}
$\rm M_*$ = 0.62 \msunm  &$\rm M_*$ = 0.56 \msunm \ (-10\%)
                                          & 16.7s & +10.1s & +18.1s & +22.0s\\
~Salaris profiles    & Basic profiles     & 15.2s & -7.40s & -32.3s & -6.01s\\
~\teffm = 12400 K& \teffm = 11160 K (-10\%)&12.9s & +12.1s & +15.4s & +11.3s\\
~$\rm M_H=10^{-7.7}$ & $\rm M_H=10^{-6.9}$ (-10\% in log) 
                                          & 4.59s & -0.273s& +3.29s & +10.2s\\
\rule[-0.3cm]{0mm}{0.6cm}
$\rm M_{He}=10^{-2.3}$& $\rm M_{He}=10^{-2.1}$ (-10\% in log)
                                          & 3.88s & -5.19s & +1.95s & -4.50s\\
\hline
\rule[-0.0cm]{0mm}{0.6cm}
$\rm X_o$=0.76 & $\rm X_o$=0.68 (-10\%)   & 3.15s & +0.58s & +7.18s & +1.69s\\
Soft C/He transition  & Sharper C/He transition  
                                          & 2.61s & -4.28s & +0.713s& -2.84s\\
\rule[-0.3cm]{0mm}{0.6cm}
$\rm q_{fm}$=0.50   & $\rm q_{fm}$=0.45 (-10\%) 
                                          & 1.18s &+0.0439s& -2.70s &-0.791s\\
\hline
\rule[-0.0cm]{0mm}{0.6cm}
\citet{enveos}       & \citet{oldeos}     & 1.08s &       &        &      \\
~envelope EOS        & envelope EOS       &               &        &      \\
~$\alpha = 0.6$      & $\alpha = 2 $      & 0.0104s       &        &      \\
\rule[-0.3cm]{0mm}{0.6cm}
ML2                  & ML1                & 0.0000689s   &        &      \\
\hline
\hline
\end{tabular}
\end{center}
\end{table}

\begin{figure}[!ht]
  \centering{\includegraphics[width=0.8\textwidth]{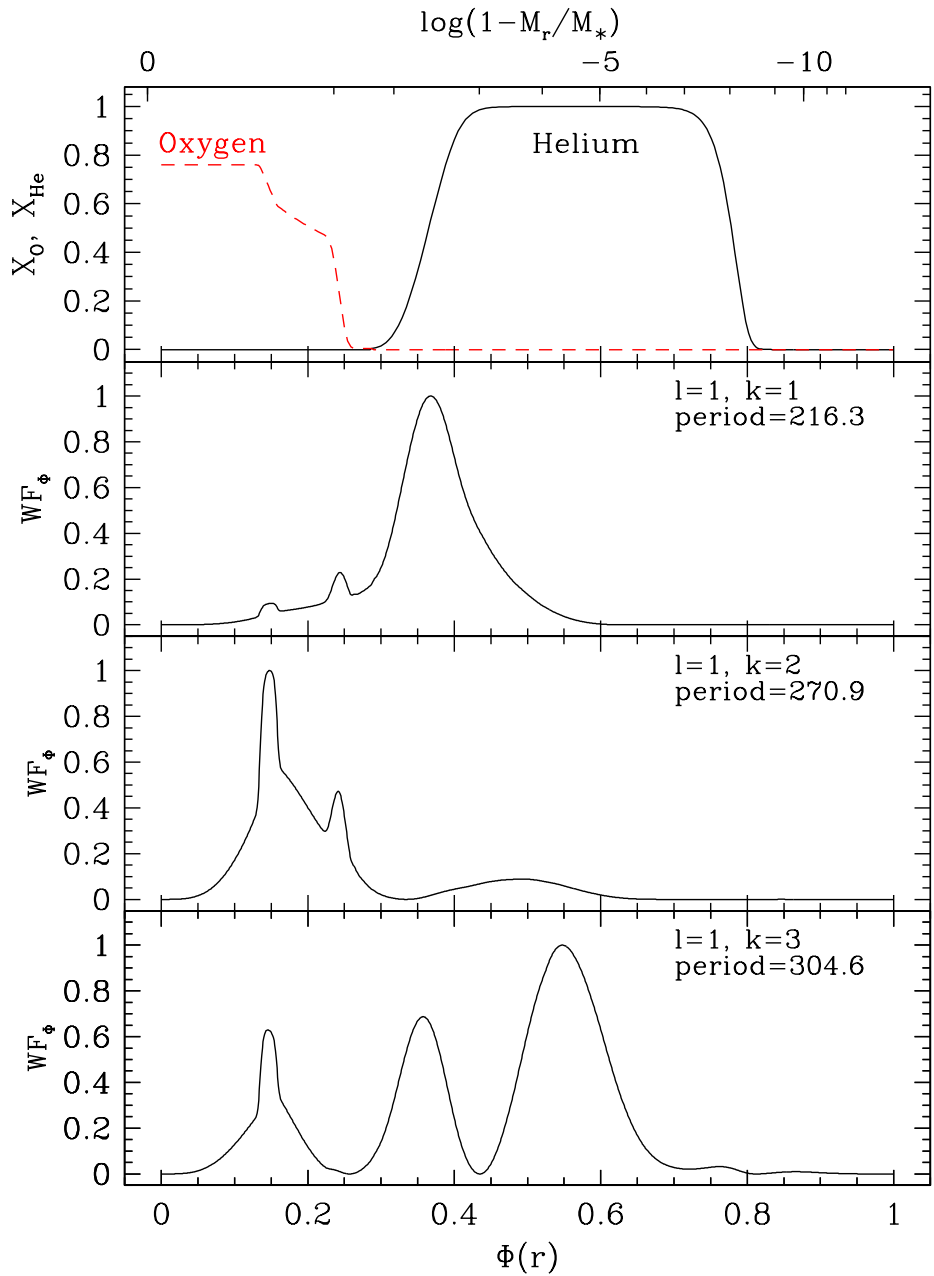}}
  \caption{
    \label{wfunc}
    A plot of the weight functions of the first three $\ell=1$
    overtones as a function of the coordinate $\Phi(r)$, the
    ``normalized buoyancy radius'', for the fiducial model of
    Table~\ref{uncertainties}.  We see that the $k=1$ mode has its
    period mainly determined by the composition gradient at the base
    of the He layer, and that the $k=2$ mode is most sensitive to the
    structure in the C/O profile in the core. The $k=3$ mode is
    sensitive to multiple features in the model and the only one
    sensitive to the location of the base of the hydrogen layer.}
\end{figure}

Examination of how each change affects each mode individually reveals a few
striking features. For instance, the 271s mode is strongly affected by a change
in the core parameter $X_o$ ($\Delta P_{271s}=7.18$s), while the other two 
modes are not ($\Delta P_{216s}=0.58$s and $\Delta P_{305s}=1.69$s). This, 
and other features apparent in the last three columns of table 5 can be 
explained by examining the weight functions of the relevant modes. 
\citet{mike03} used them as a diagnostic of the effect of composition 
transition zones on mode trapping. In figure~\ref{wfunc}, we show the weight 
functions of the first three $\ell=1$ modes for the fiducial model of
Table~\ref{uncertainties}. In the second panel of figure~\ref{wfunc}, we see 
that the $k=1$ mode resonates strongly with the base of the He layer. This 
shows that it is extremely sensitive to the He layer mass, and very insensitive
to the H layer mass, which is what we see from $\Delta P_{216s}$ in 
Table~\ref{uncertainties}.  Looking at the third panel, we notice that the 
$k=2$ mode predominantly samples the chemical profile in the core; its
period change due to a change in C/O profile is 5 times that of the
other two modes. This explains why the 271 s mode is so sensitive to the 
core parameters. The last panels shows that the $k=3$ mode samples both of 
these features, as well as farther out in the model near the base of the H
layer. This mode is the most sensitive of the three to the H layer
mass.

In order to keep our model grids manageable, we decided to vary 4 parameters.
While  \teffm, $\rm M_*$, $\rm M_H$, and $\rm M_{He}$ have by far the largest 
effect on the periods, the parameters associated with the core abundance 
profiles and the shape of the carbon-helium transition zone (second block in 
Table \ref{uncertainties}) can have a significant effect (as high as $\sim 3$ 
seconds) on the pulsation periods. Since we had to fix our core composition 
profiles, we decided to fix them to profiles predicted by stellar evolution
(figure \ref{fig2}). We also adopted the fiducial model's carbon-helium 
transition profiles.

In the asteroseismological fits detailed in section 5, we use the periods 
rounded to the second, and do not distinguish a model period that fit within 
0.5 seconds from one that fits within 1 second. This allows us not to worry
ourselves with the factors listed in the last block of Table 
\ref{uncertainties} and other possible small effects.

\subsection{Fine Grid Search}

We started by building a low resolution grid that covered a broad region of 
parameter space, guided by the spectroscopy (Table \ref{starsproperties}). We
varied masses between 0.46 \msunm \ and 0.80 \msunm \ and temperatures between
10800 K and 13000 K. We considered $-3 < \log(\rm(M_{He}) < -2$ and 
$-8.4 < \log(\rm M_H) < \log(\rm M_{He}) - 2$. For that grid, we determined 
that step sizes of 200K in \teffm, 0.02 \msunm \ in $\rm M_*$, and 0.4 in $\rm 
log(M_{He})$ and $\rm log(M_H)$ were sufficient to locate likely minima of 
$\Phi$. Using the results of the broad grid search, we narrowed down our search
to smaller areas of parameter space and built a finer grid. We found that a 
resolution of 100K in \teffm, 0.01 \msunm \ in $\rm M_*$, and 0.2 in $\rm 
log(M_{He})$ and $\rm log(M_H)$ was sufficient to clearly define the minima of 
$\Phi$. The broad grid contains near 20000 models, the fine grid 6000 (it is 
more limited in parameter space). 

For G117-B15A, we further narrowed down the list of possible best fits by 
assuming that G117-B15A's modes were $\ell=1$ modes, consistent with mode 
identification work done by \citet{elr95} and the fact that higher l modes 
suffer from geometric cancellation at the surface of the star and are likely to
result in low-amplitude pulsations. For R548, we required that the two high 
amplitude modes, which also seem to be present in G117-B15A (see Table 
\ref{starsproperties}), be $\ell=1$ modes. We did not place any constraints on 
the identification of the other three modes.

\section{Results}

\subsection{Mass and Effective Temperature}

\begin{figure}[!ht]

\begin{center}
\rotatebox{270}{\scalebox{1.}{\includegraphics[width=280pt,height=200pt,bb=25 200 580 600]{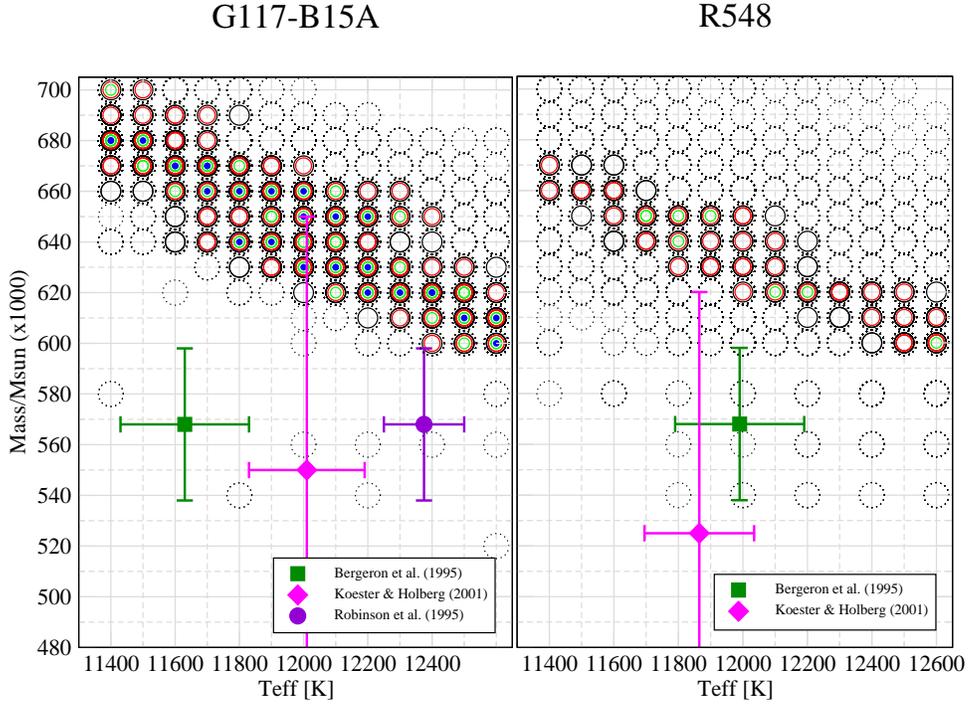}}}
\end{center}
\caption{
\label{finemassteff}
The valley of best fit models for G117-B15A and R548 in the \massteff plane. 
The dashed circles mark the location of the subset of models that fit the 
$\ell=1$ mode identification criterion (see text). Of those, the progressively 
filled-in circles indicate better and better fits ($\Phi <$ 2.5s, 2s, 1.5s, 1s 
respectively). The filled symbols with error bars indicate the 
spectroscopically determined temperatures and mass for G117-B15A and R548, 
according to the legend. We used our models to derive a mass from the gravities
listed in Table \ref{starsproperties}.
}
\end{figure}

We display the results of the fine grid search for G117-B15A and R548 side by 
side in the \massteff plane in figure \ref{finemassteff}. We explored all of
the parameter space shown. The fine grid starts at 0.6 \msunm and above. We 
also indicate the spectroscopic temperature determinations for both stars, as 
well as masses derived from the gravities listed in Table 
\ref{starsproperties}, using our models. Our best fit models appear 
systematically massive and/or hot compared to the spectroscopic results. We
shall come back to this discrepancy in section 6.

With our 1s sensitivity cut-off due to small modeling uncertainties, we are 
unable to determine a unique point in parameter space that matches G117-B15A or
R548. Instead, we find families of solutions. There is a tight correlation 
between the mass and the effective temperature, anticipated from earlier work 
\citep{paul98}. Decreasing either the temperature or the mass decreases the 
spatial average of the \bvf frequency in a similar way, and therefore yields 
similar sets of periods. In general, our best fit models are more massive 
and/or hotter than we would have expected from the spectroscopy.

From their observed properties (Table \ref{starsproperties}), we expected 
G117-B15A and R548 to be best fit by similar models. We do see that in our 
results. It also comes as no surprise that the $\ell=1$ identification 
requirement does not limit model fits to R548 as much as they do for G117-B15A.
To obtain the dotted circles in figures \ref{finemassteff}, we discarded 
fits that did not obey the $\ell=1$ mode identification for G117-B15A's 3 modes
and for R548's 2 high amplitude modes (regardless of the quality of the fit). 
Because this first cut is based on the constraint of a fewer number of modes in
the case of R548, it does not eliminate as many models. On the other hand, we 
have 3 additional modes to fit for R548 and they ultimately allow us to narrow 
down the best fit models to a small region of parameter space. We do not have 
that luxury with G117-B15A, as the 3 $\ell=1$ modes are all we have. We also 
obtain better fits to G117-B15A's 3 periods than we do to R548's 5 periods, as 
we would expect.

\subsection{Helium and Hydrogen Layer Mass}

For both stars, the helium layer mass appears to be fairly well
constrained around $4 \times 10^{-3}$. It is determined by the 215s
mode in G117-B15A and the 213s mode in R548. If we do not include
those modes in the fit, $\rm M_{He}$ is essentially unconstrained. For
G117-B15A, if we change that mode by as little as 5 seconds (e.g. 215s
to 210s), the helium layer mass changes from $4 \times 10^{-3}$ to $5
\times 10^{-3}$. If instead we leave out the 271s mode, the helium
layer mass remains constrained. This result is readily understandable
from figure~\ref{wfunc} and the associated discussion of weight
functions in section~4.2.

\begin{figure}[!ht]
\begin{center}
\rotatebox{270}{\scalebox{1.}{\includegraphics[width=290pt,height=200pt,bb=15 200 580 600]{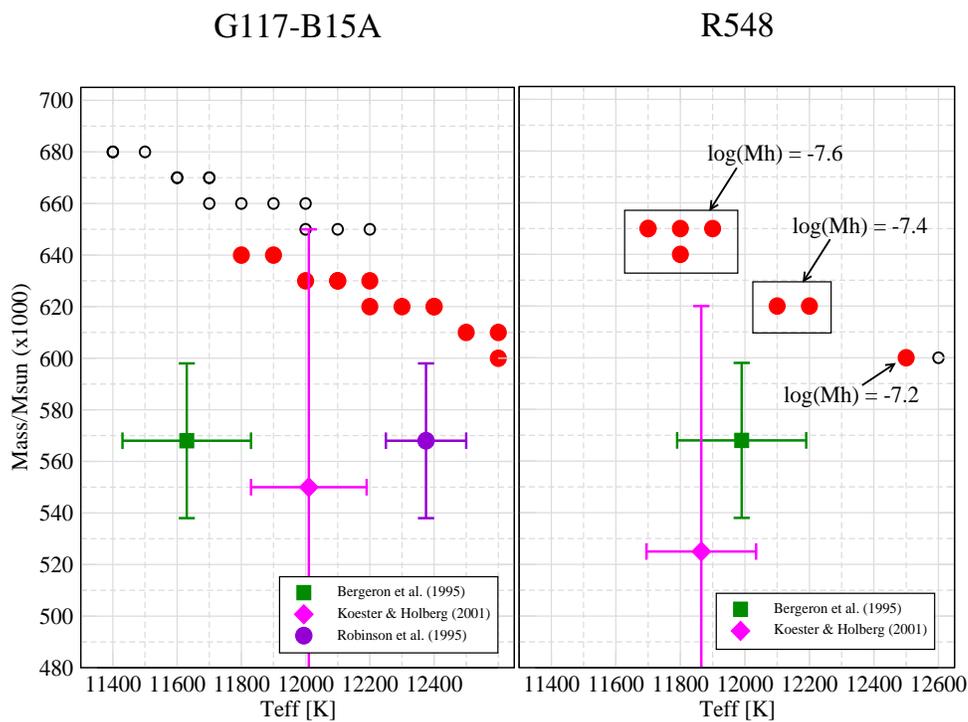}}}
\end{center}
\caption{
\label{finebymh}
Best fit models in the \massteff plane for different hydrogen layer masses. For
G117-B15A, we isolated the solutions for which $\Phi < 1$s and for R548, those
for which $\Phi < 1.5$s. The open circles correspond to ``thick'' hydrogen 
solutions ($\rm M_H = 6.3 \times 10^{-7}$), while the filled circles correspond
to thin hydrogen solutions ($\rm M_H \simeq 1-4 \times 10^{-8}$).
}
\end{figure}

In figure \ref{finebymh}, we show the very best fit models ($\Phi \leq 1$s for
G117-B15A and $\Phi \leq 1.5$s for R548) in the \massteff plane and indicate 
their respective hydrogen layer masses. We discover that families of models 
with different hydrogen layer masses separate out in the \massteff plane.

We find that R548 is best fit with thin hydrogen layer models 
($\rm M_H \simeq 2-6 \times 10^{-8}$). For G117-B15A, we find two well
defined families, one at higher mass and lower effective temperature with 
$\rm M_H = 6.3 \times 10^{-7}$ and one at lower mass and higher effective 
temperature with $\rm M_H \simeq 1-4 \times 10^{-8}$. Previous investigations 
have found both thick and thin solutions \citep{paul98,benvenuto}. Based on the
incompleteness of our current understanding of mass loss in the late stages of
stellar evolution, we do not believe either set of models can be ruled out by 
stellar evolution calculations. For instance, \citet{althaus02} present
the hydrogen abundance ($M_H = 10^{-4}$) in DAVs as an upper limit, as it would
be reduced by the inclusion of mass loss episodes during the planetary phase, 
the extent of which is unknown. While none of our best fits truly have thick 
hydrogen layers, we shall refer to the two distinct families as ``thick'' and 
``thin'' hydrogen layer fits.

\subsection{Mode Identification}

Previous asteroseismological studies of G117-B15A \citep{paul98,benvenuto} 
identified the three modes as consecutive $\ell=1$ modes, with the 215s mode
being either a $k=1$ or a $k=2$ mode. \citeauthor{paul98} found equally good 
fits with either mode identification. While \citeauthor{benvenuto}'s best fit 
model was consistent with a $k=1,2,3$ mode identification, they also found good
fit models with $k=2,3,4$ modes. Among the 27 models in the fine grid that 
matched the observed periods to better than 1 second on the average, we found
13 matched the observed periods with $k=1,2,3$ modes and 14 with $k=2,3,4$ 
modes. The former family of models all have $\rm M_H \simeq 1-4 \times 10^{-8}$
and the latter $\rm M_H = 6.3 \times 10^{-7}$. This is qualitatively in tune
with what \citet{paul98} and \citet{benvenuto} found (the $k=1,2,3$ fits have 
thin hydrogen layers and the $k=2$,3,4 fits have thicker hydrogen layers).

For R548, among the top 20 best fit models, we find that the most likely mode 
identification (17 out of 20 models) is $\ell=1$, $k=1,2,4$ respectively for 
the 213s, 274.5, and 318s modes and $\ell=2$, $k=4$ and 8 for the 187s and 334s
modes. Recall that we required the 213s and 274.5s modes to be $\ell=1$ modes, 
but did not place any constraints on the other modes. Two out of the three 
models that disagree with this mode identification are thick hydrogen models 
(while the 17 models with the most common mode identification are thin hydrogen
models). The $\ell=2$, $k=4$ identification for the 187s mode is very robust 
(20/20).

The main reason we are studying G117-B15A and R548 side by side is that they 
are observationally similar and we therefore expect them to also be 
structurally similar. If that is the case, then the mode identification results
for R548 suggest that the correct mode identification for G117-B15A is $k=1$,2 
and 3, in favor of the thin hydrogen layer and lower stellar mass solutions 
($\rm M_H \simeq 1-4 \times 10^{-8}$). 

\section{Summary and conclusions}

We performed a systematic fine grid search for best fit models to the two
DAVs G117-B15A and R548. We find best fit models for both stars between 11600 
and 12700 K and between 0.59 and 0.66 \msunm. In both cases, the region 
occupied by the best fit models in the \massteff plane is related to the 
thickness of the hydrogen layer. Treating the helium layer mass as a free 
parameter, we discovered that the lowest period mode for each star (215s for 
G117-B15A and 213s for R548) singlehandedly sets the helium layer mass to 
$4 \times 10^{-3}$. Both stars are well fit with thin hydrogen layer models 
($\rm 10^{-7} < M_H < 10^{-8}$). For G117-B15A, we find a second family of 
solutions between 11300 and 12300 K and between 0.65 and 0.68 \msunm. Those 
fits have slightly thicker hydrogen layers ($\rm M_H = 6.3 \times 10^{-7}$).

For R548 we find a unique, robust mode identification. The dominant modes 
(213s and 274.5s) are $\ell=1$, $k=1,2$ modes. The 318s mode is also an 
$\ell=1$ mode, with $k=4$. The last two modes (187s and 334s) are $\ell=2$, 
$k=4$ and 8 respectively. For G117-B15A, we find two distinct families of best 
fit models. Models with $\rm M_H = 6.3 \times 10^{-7}$ all have the same mode 
identification, namely $k=2,3$, and 4 for the three observed periods 
(215s, 271s, and 304s). Models with $\rm M_H \simeq 1-4 \times 10^{-8}$ are 
consistent with G117-B15A's periods being $k=1,2$, and 3. R548 and G117-B15A 
have similar observed properties and based on that fact, it is likely that 
they have similar structures. In this case, the second class of models appears 
better, but we cannot discard the first class of models based on that fact 
alone.

By sampling parameter space systematically and homogeneously, we found
that our models were offset in mass and temperature compared to the
spectroscopy (the models are hotter and/or more massive).
\citeauthor{reid96} (1996) measured the gravitational redshift of spectral
lines for G117-B15A. The resulting mass is in tune with the
spectroscopic mass ($0.536 \pm 0.010 \msun$). Both different equation of state 
tables and opacities could alter the structure of the models and possibly 
modify our results. We do not expect either to have a large effect, however. 
Just recently, \citet{cassisi} published new electron-conduction opacities that
treat the partially degenerate regime relevant in white dwarf envelopes better.
According to \citeauthor{cassisi}, the new opacities differ only by at most a 
factor of 2 in white dwarf envelopes and have very little effect on the 
mass-radius relation. We also tried to revert back to the \citet{oldeos} 
envelope equations of state and found that this had a negligible effect on the 
gravities in our models.

Results from section 4.2 suggest a more promising avenue. They show that 
structure in the core and the shape of the helium composition profile at the 
carbon-helium interface can have a large effect on the pulsations periods of a 
model. We adopted core composition profiles from stellar evolution calculations
and did not vary them. Our asteroseismological results suggest that perhaps we 
should try different core composition profiles. With precise enough 
determinations of G117-B15A and R548's mass and effective temperature from 
spectroscopy or other independent methods, one can turn the problem around 
and attempt to determine what composition profiles are needed in order to 
reconcile the asteroseismology with the spectroscopy. While preliminary 
results appear promising, a full investigation requires the clever analysis of 
a more extensive grid of models and we leave that for a future publication.

Once we are satisfied that we have models that match both asteroseismologically
and spectroscopically, we can calculate \pdots \ for our models and use them 
together with the observed \pdots \ to constrain the energy loss rate due to 
any weakly interacting particles, such as axions. The fine grid approach allows
us to formaly assess the uncertainties in our models' parameters and to obtain
tight constraints on the emission rate of weakly interacting particles in 
G117-B15A.

\section{Acknowledgments}

We wish to thank our referee for helping us improve on our work and giving us
new ideas. We thank Dr. E. Robinson for providing essential feedback. This work
was made faster and easier thanks to neatly packaged code provided by Dr. T. 
Metcalfe. This research was supported by NSF grant AST-0507639. 

\bibliographystyle{apj}

\clearpage


\end{document}